\documentclass[traditabstract]{aa}

\usepackage{epsfig}
\usepackage[switch]{lineno}
\usepackage{graphicx}
\usepackage[varg]{txfonts}
\usepackage{natbib}
\usepackage{url}
\usepackage{ulem}

\bibpunct{(}{)}{;}{a}{}{,} 

\def\ecs{erg~cm$^{-2}$s$^{-1}$}
\def\lum{erg~s$^{-1}$}
\def\bron{SAX J1808.4$-$3658}
\def\mjdstart{55873.91320}

\begin{document}

\title{A bright thermonuclear X-ray burst\\ simultaneously observed
  with {\it Chandra}\/ and {\it RXTE}} 
 \titlerunning{A bright thermonuclear X-ray burst observed with {\it
 Chandra}\/ and {\it RXTE}}

\authorrunning{in 't Zand, Galloway, Marshall, Ballantyne et al.}

\author{J. J. M.~in~'t~Zand\inst{1}, D.K. Galloway\inst{2},
  H.L. Marshall\inst{3}, D.R. Ballantyne\inst{4},
  P.G. Jonker\inst{1,5,6}, F.B.S. Paerels\inst{7},
  \\ D.M. Palmer\inst{8}, A. Patruno\inst{9} \& N.N. Weinberg\inst{3}}


\institute{     SRON Netherlands Institute for Space Research, Sorbonnelaan 2,
                3584 CA Utrecht, the Netherlands; {\tt jeanz@sron.nl}
           \and
                Monash Centre for Astrophysics, School of Mathematical
                Sciences \&  School of Physics, Monash University, VIC 3800,
                Australia
           \and
                Dept. Physics and Kavli Institute for Astrophysics and Space
                Research, Massachusetts Institute of Technology, Cambridge,
                MA 02139, U.S.A.
           \and
                Center for Relativistic Astrophysics, School of Physics,
                Georgia Institute of Technology, Atlanta, GA 30332, USA
           \and
                Harvard-Smithsonian Center for Astrophysics, 60 Garden Street,
                Cambridge, MA 02138, U.S.A.
           \and
                Department of Astrophysics/IMAPP, Radboud University Nijmegen,
                PO Box 9010, 6500 GL Nijmegen, the Netherlands
           \and
                Columbia Astrophysics Laboratory, 550 West 120th Street,
                New York, NY 10027, U.S.A.
           \and
                Los Alamos National Laboratory, B244, Los Alamos, NM 87545,
                U.S.A.
           \and
                Astronomical Institute 'Anton Pannekoek', University of
                Amsterdam, Science Park 904, 1098 XH Amsterdam, The Netherlands
          }

\date{\it Accepted for publication March 19, 2013}

\abstract{The prototypical accretion-powered millisecond pulsar
  \bron\ was observed simultaneously with {\it Chandra}-LETGS and {\it
    RXTE}-PCA near the peak of a transient outburst in November 2011.
  A single thermonuclear (type-I) burst was detected, the brightest
  yet observed by {\it Chandra} from any source, and the
  second-brightest observed by {\it RXTE}.  We found no evidence for
  discrete spectral features during the burst; absorption edges have
  been predicted to be present in such bursts, but may require a
  greater degree of photospheric expansion than the rather moderate
  expansion seen in this event (a factor of a few). These observations
  provide a unique data set to study an X-ray burst over a broad
  bandpass and at high spectral resolution ($\lambda/\Delta
  \lambda=200$--400).  We find a significant excess of photons at high
  and low energies compared to the standard black body spectrum. This
  excess is well described by a 20-fold increase of the persistent
  flux during the burst. We speculate that this results from burst
  photons being scattered in the accretion disk corona. These and
  other recent observations of X-ray bursts point out the need for
  detailed theoretical modeling of the radiative and hydrodynamical
  interaction between thermonuclear X-ray bursts and accretion disks.

}

\maketitle

\section{Introduction}
\label{secintro}

The vast majority of thermonuclear X-ray bursts from neutron stars
\citep[the so-called type I X-ray bursts, see e.g.][]{lew93,stroh06}
have been measured with instruments that are sensitive at photon
energies above 2 keV. The temperature of type-I X-ray bursts is,
however, 2.5 keV or lower so that usually a substantial part of the
spectrum is missed.  Therefore, interesting physics may be missed,
particularly during the cooler parts of X-ray bursts in the tail and
during phases of photospheric expansion. Not only that: many of the
discrete atomic spectral features from abundant elements in the cosmos
(oxygen, neon and iron) occur below 2 keV.  Sub-2 keV measurements of
X-ray bursts at high spectral resolution and with photospheric
expansion are, therefore, interesting to pursue.

Since the launch of {\it Chandra}\/ and {\it XMM-Newton}\/ in 1999 and
{\it Swift}\/ in 2004, sub 2 keV coverage is readily
available. However, these are narrow field instruments where
measurements are usually made through a dedicated program, in contrast
to the wide-field instruments on board SAS-B, BeppoSAX, INTEGRAL, {\it
  Swift}\/ and {\it Fermi}\/ that lack sub 2 keV coverage. We estimate
that about 250 X-ray bursts have been detected below 2 keV with these
three missions, compared to about 10$^4$ with the wide field
instruments
\citep[e.g.,][]{cot02,jon03,tho05,boi07,kon07,pai12}. Hardly any X-ray
bursts with photospheric expansion and low $N_{\rm H}$ were detected
with sub 2 keV coverage so far, perhaps 6 with low flux \citep{gal10}.

In this paper we report an observation of a bright thermonuclear burst
with photospheric expansion from the low-mass X-ray binary (LMXB)
\bron, with the AXAF CCD Imaging Spectrometer (ACIS) in combination
with the Low-Energy Transmission Grating Spectrometer (LETGS) on board
{\it Chandra}.  The event was detected simultaneously with the {\it
  RXTE}\/ Proportional Counter Array (PCA). Combined this presents a
unique data set with, as far as we know, the largest number of photons
ever detected from a single X-ray burst in the 0.5--2~keV band or with
a grating spectrometer.

\bron\ is the first discovered accretion-powered millisecond X-ray
pulsar in a LMXB \citep{zan98,wijn98,zan01,chak03}, of 14 cases known
thus far \citep[e.g.,][]{pat12a}. It is also a relatively nearby
representative, at $3.5\pm0.1$ kpc \citep{gal06}, and with a
relatively low column density of interstellar material \citep[$N_{\rm
    H}=(1-2)\times10^{21}$~cm$^{-2}$;
  e.g.][]{wan06,pap09,cac09}. Almost all its X-ray bursts exhibit
photospheric expansion \citep{gal08} and the bursts appear to be the
result of a flash of a pure helium layer that is produced by stable
hydrogen burning. These circumstances make \bron\ a particularly
suitable source to study X-ray bursts in great detail and in a large
bandpass. Therefore, we observed \bron\ with {\it Chandra}\/ and {\it
 RXTE}\/ during its November 2011 outburst.

\begin{figure}[t]
\vspace{-1.5cm}
\includegraphics[width=\columnwidth,angle=0]{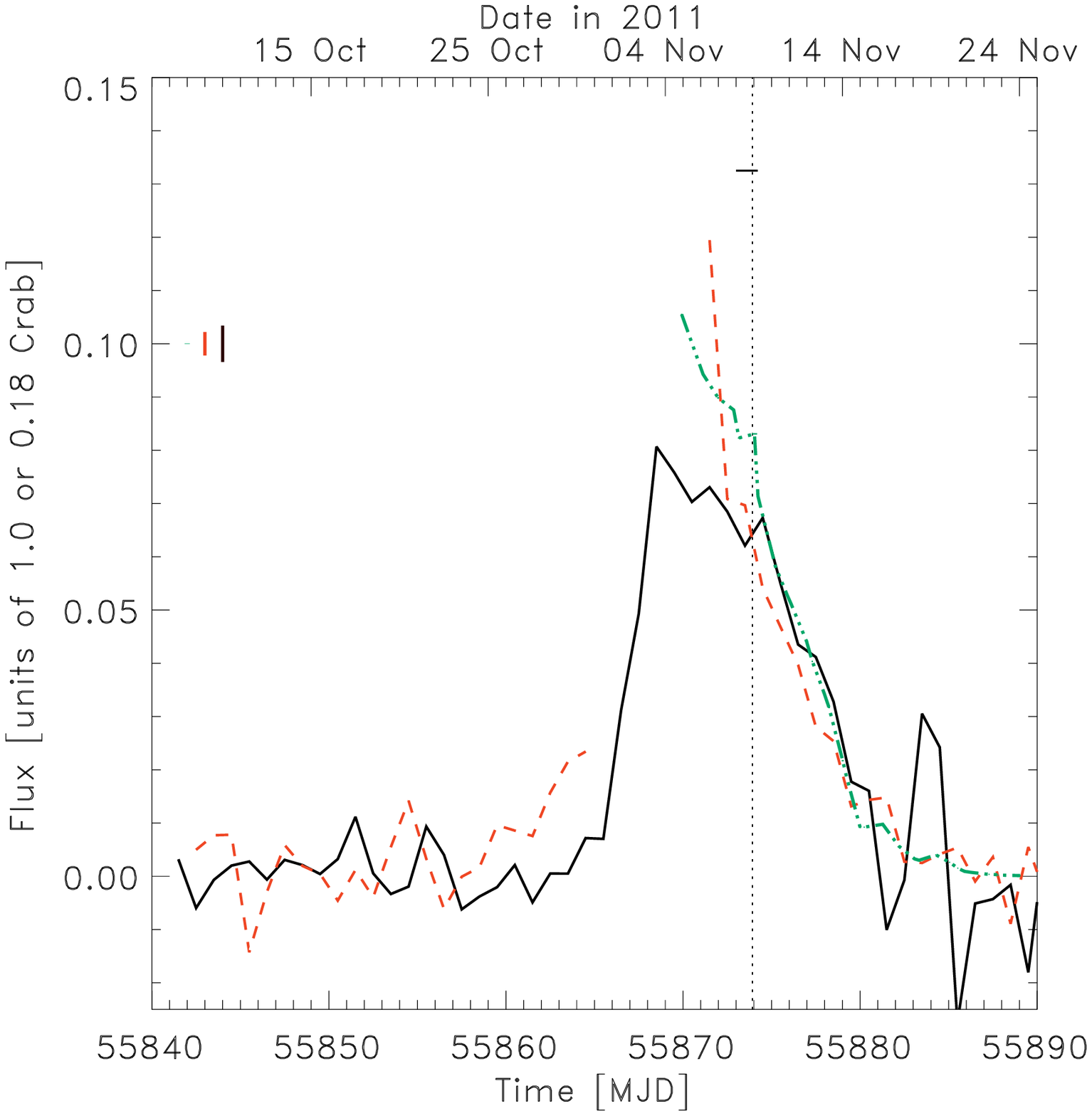}
\vspace{-3.5cm}
\caption{Light curves of the 2011 outburst of \bron: from 2--10~keV
  {\it RXTE}/PCA data \citep[green dash-dotted line, normalized to
    1580.5~c~s$^{-1}$PCU$^{-1}$ equivalent to 1.0 Crab unit; see
    also][]{pat12b}, 2--10~keV {\it MAXI}/GSC data (red dashed line;
  normalized to 3.2~c~s$^{-1}$cm$^{-2}$ equivalent to 1.0 Crab unit)
  and 15--50~keV {\it Swift}/BAT data (black solid line; normalized to
  0.22~c~s$^{-1}$cm$^{-2}$ equivalent to 0.18 Crab units). The
  outburst started on MJD~55865 (Oct. 31st) and lasted for 21~d, until
  MJD~55886. The horizontal solid line indicates the time frame of the
  {\it Chandra}\/ observation and the vertical dotted line through it
  the burst that was detected with {\it Chandra} and {\it RXTE}. The
  vertical bars at mid left indicate the typical 1$\sigma$ uncertainty
  per data set (PCA flux errors are typically smaller than the line
  thickness).
\label{figoutburst}}
\end{figure}

\begin{figure}[t]
\includegraphics[width=\columnwidth,angle=0]{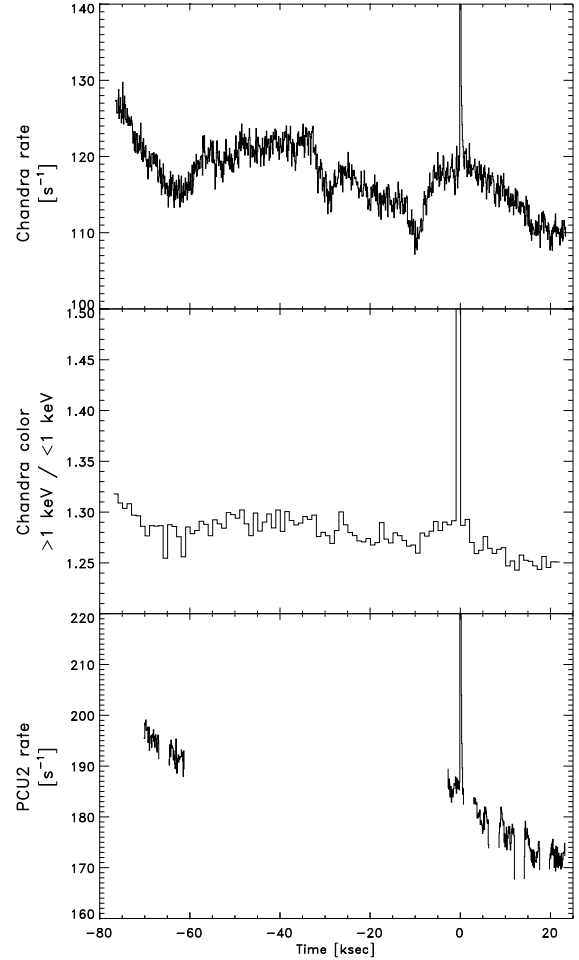}
\caption{Time histories of total {\it Chandra}-observed photon rate,
  measured over all active ACIS-S CCDs, at 100 s resolution (top
  panel), the color at 1000 s resolution (middle panel) and the {\it
    RXTE}\/ PCU2 rate at 100 s resolution (bottom panel). The burst
  peak has been cut in all panels. No background was subtracted.}
\label{figlc}
\end{figure}

The main goal of our measurement was to search for absorption
edges. We were motivated by a prediction of \cite{wei06} that the
radiative wind of radius expansion bursts eject ashes of nuclear
burning whose spectral signature may be detectable with current
high-resolution X-ray telescopes.  This may be the best path to
finding discrete spectral features from neutron star (NS) surfaces,
whose detection could constrain thermonuclear reactions in the NS
ocean and the NS compactness through general relativistic
gravitational redshift determinations. Near-Eddington X-ray bursts
reveal themselves as having black body like spectra with phases of
expanded emission areas. There are two kinds: those with moderate
expansion (factor $\sim10$ increase in emission area), encompassing
$\approx20$\% of all bursts, and those with superexpansion (factor
$\ga10^4$), encompassing perhaps $\sim0.1$\% of all bursts
\cite[e.g.][]{jvp90,gal08,zan10}. The difference in expansion is
likely determined by whether an optically thick shell is being
expelled which on its turn is determined by how much the radiative
flux from the nuclear burning goes over the Eddington limit in the
flash layer. Also, the larger this super-Eddington factor is, the
larger the upward extent of the convection and the probability for
absorption edges will be. The burst we discuss here is a bright burst
from a relatively nearby NS and, ergo, of good statistical
quality. However, it has only a moderate photospheric expansion.

We introduce the observations in Sect.~\ref{secobs}, including
monitoring observations with the Monitoring All-sky X-ray Imager ({\it
  MAXI}) on board the International Space Station (ISS) and the Burst
Alert Telescope (BAT) on board {\it Swift}, and provide the details of
the data reduction in Sect.~\ref{secred}. Sect.~\ref{secspec}
discusses the spectral analysis and Sect.~\ref{sectiming} briefly the
timing behavior of the detected burst.  Finally, in
Sect.~\ref{secdiscuss}, we place the results in the context of the
theory of burst spectra.

\section{Observations}
\label{secobs}

\begin{figure}[t]
\vspace{-0.cm}
\includegraphics[width=\columnwidth,angle=0]{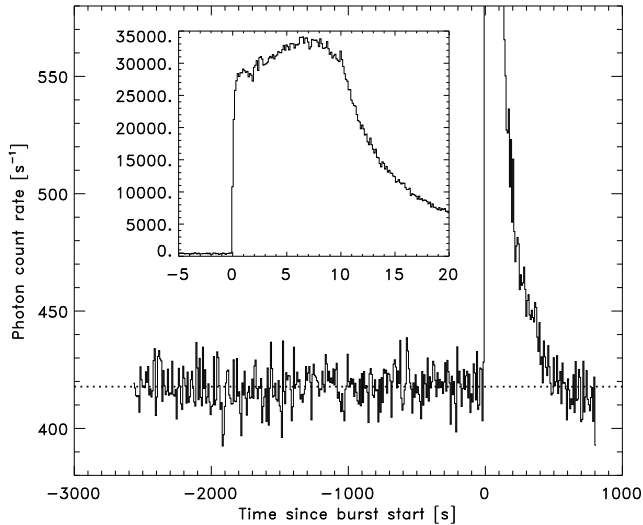}
\vspace{-0cm}
\caption{Time history of {\it RXTE}\/ detected photon rate around the
  burst, at 8 s resolution. Photons detected with PCUs 1 and 2 have
  been combined. The dotted horizontal line shows the fit with a
  constant line to the data prior to the burst (with a best-fit value
  of 418 s$^{-1}$). The burst lasts $550\pm20$~s. The inset zooms in
  on a 25 s stretch around the peak, at 1/8 s resolution.}
\label{figlcxte}
\end{figure}

On November 4, 2011, a new outburst of \bron\ was reported from {\it
  Swift}-BAT observations beginning October 31 \citep{mar11}.
Fig.~\ref{figoutburst} shows the outburst light curves from {\it
  MAXI}, {\it RXTE}-PCA and {\it Swift}-BAT data. {\it MAXI}
\citep{mat09} is deployed on the International Space Station (ISS)
and, in combination with the Gas Scintillator Counter (GSC), is active
between 2 and 30 keV. The measurements consist of 1~min exposures
every 100-min ISS orbit and are made public within a few hours of
being taken\footnote{\url{http://maxi.riken.jp/top}}. BAT
\citep{bar05} is a gamma-ray burst monitor camera active on {\it
  Swift}\/ between 15 and 100 keV. The 15-50 keV flux of some 500
X-ray sources are monitored several times per day and made available
publicly in near real
time\footnote{\url{http://swift.gsfc.nasa.gov/docs/swift/results/transients}}. The
PCA data (see below) were obtained under ObsID 96027-01 for a total
exposure time of 143.6~ks \citep[see also][and below]{pat12b}.

Following the alert, we requested and were granted a {\it Chandra}\/
Target of Opportunity Observation (ObsID 13718). The exposure time is
100.13 ks between Nov. 8 00:46 UT and Nov. 9 04:35 UT (see
Fig.~\ref{figlc}) which is 8 days after the presumed outburst
onset. The observatory was operated using the Low-Energy Transmission
Grating Spectrometer \citep[LETGS;][]{bri00}, with ACIS-S
\citep{gar03} at the focal plane and the Low-Energy Transmission
Grating in the beam. This combination provides a marginally higher
effective area than the HETGS (see e.g.  Proposers' Observatory Guide
Fig. 1.4) in most of the interesting bandpass of 0.5-6 keV at the
expense of some spectral resolution. The resolution is, however,
sufficient to search for absorption edges ($\lambda/\Delta
\lambda=200$--400). A disadvantage of using ACIS instead of HRC is the
loss of effective area between 0.08 and 0.5 keV, but this bandpass is
of no interest to this observation because the source is significantly
absorbed there. To avoid pile up, ACIS was operated in the continuous
clocking (CC) mode. In this mode, position information is lost in the
cross dispersion direction while the time resolution is improved from
3.2~s to 2.85~ms. The disadvantage is that the background cannot
accurately be measured independently (it is possible to discriminate
with some sensitivity the background from source in pulse height
versus dispersion space), but this is not detrimental because we are
primarily interested in X-ray bursts which can easily be separated in
the time domain, without the need for the spatial resolution. Four
ACIS CCDs were employed: S1 through S4. The offset of the Science
Instrument Module (the so called 'SIM offset') was -8 mm to position
the dispersed spectrum as close as possible to the CCD edges to
optimize the CCD spectral resolution.  The Y offset was +1.6 arcmin so
that the zeroth order would be detected by S2.

Simultaneous measurements were requested and granted with {\it RXTE}\/
under Proposal Number 96027. Twenty-four ks of simultaneous coverage
was obtained (see Fig.~\ref{figlc}). The Proportional Counter Array
\citep[PCA;][]{jah06} comprises 5 non-imaging proportional counter
units (PCUs) active between 2 and 60 keV with a combined effective
area of 6000~cm$^2$ at 6 keV. The spectral resolution is 18\% full
width at half maximum at 6 keV and the time resolution of the data
products for \bron\ is typically 125 $\mu$s. It is seldom that all
PCUs are active at the same time, particularly at late times in the
mission. During our observations, PCA operated with 2 active
proportional counter units (PCUs 1 and 2, counted from 0). All PCUs
have collimators delimiting the field of view to
2\degr$\times$2\degr\ (full width to zero response). For our analysis,
we used standard-1 and 2 data (time resolution 0.125 and 16 s, energy
resolution 1 and 128 channels, respectively) and event mode data {\tt
  E\_125us\_64M\_0\_1s} with 125 $\mu$s time resolution and 64 energy
channels between 2 and 60 keV. All data allow selection of PCUs.

Fig.~\ref{figlc} shows the {\it Chandra}\/ and PCA-measured (only
PCU2) light curves during the time of the {\it Chandra}\/
observation. These curves show the detection of one burst in both
instruments. This is the only X-ray burst detected from \bron\ in the
whole outburst \citep[c.f.,][]{pat12b}.  {\it RXTE}\/ and {\it
  Chandra}\/ times were synchronized by matching the burst onset as
seen in the overlapping bandpass 3-8 keV.  We find that the burst
starts at MJD~\mjdstart\ (barycentered), or at {\it RXTE}\/
time\footnote{Mission Elapsed Time, defined as seconds since
  1994.0~UTC} 563407175.90~s and {\it Chandra}\/ time\footnote{
  Mission Elapsed Time, defined as seconds since 1998.0 TT}
437176836.46~s (not barycentered).  There is simultaneous {\it
  Chandra}/{\it RXTE}\/ coverage between -2560 and +812 s with respect
to the burst onset time (see Fig.~\ref{figlcxte}). The total number of
burst photons detected is 38,800$\pm$180 with {\it Chandra}\/
(measured over the complete ACIS-S detector and after subtraction of a
pre-burst level) and 562,800$\pm$400 with {\it RXTE}. The peak
intensities, measured at 0.125~s time resolution, are 2,538 c~s$^{-1}$
for {\it Chandra}\/ (all ACIS events, not corrected for non-burst
events) and 33,669 c~s$^{-1}$ for {\it RXTE}. The {\it Chandra}\/ net
peak rate for orders -1/+1 is 1,930 c~s$^{-1}$. The {\it Chandra}\/
event data buffer of 128,000 events (in the graded telemetry mode) was
not overfilled. The {\it RXTE}\/ data buffer was overfilled. Starting
at 1.15 s after burst onset, 0.45 s chunks of data are missing in the
event-mode data every 1.0 s until 10.6 s. After that 3 smaller chunks
of 0.45 s are missing every 1.0 s until 13.6 s after which the data
are completely recovered. The standard-1 and 2 data do not have gaps.

The non-burst emission varied significantly during the 100~ks of the
{\it Chandra}\/ observation. The intensity decreased from 128 to 110
c~s$^{-1}$, or 14\% although this is not corrected for the background
(which is expected to be a few c~s$^{-1}$). The decrease is not
smooth. The {\it Chandra}\/ color history, the color being defined as
the intensity of all ACIS events above 1 keV divided by that below,
also shows a gradual decrease by 5\%.

\section{Data reduction}
\label{secred}

There are 11.79 million events in the {\it Chandra}\/ event file,
equivalent to an average count rate of 117 c~s$^{-1}$, in 4 ACIS-S
detectors. The 0th order contains 3.56 million events (30\%), -1/+1
orders contain 2.43/3.29 million events, orders -2/+2 0.12/0.14
million events, orders -3/+3 0.27/0.28 million events. We investigated
only orders -1 and +1 since these encompassed 7.1 times more photons
than orders -2/+2 and -3/+3 combined.

We reduced the {\it Chandra}\/ data with {\tt ciao} version 4.4, {\tt
  ftools} version 6.12 and CALDB version 4.5.0. {\tt tgextract} was
used to extract the grating order data from the event-2 file, after
the eventfile was filtered for a particular time interval with {\tt
  dmcopy}, including a
workaround\footnote{\url{http://cxc.harvard.edu}} for a bug in {\tt
  ciao} versions 4.3 and 4.4 concerning incorrect GTIs when
time-filtering grating data with {\tt dmcopy}. The resulting pha2 file
was divided into separate orders with {\tt dmtype2split}.  {\tt
  fullgarf} was used to obtain ancillary files for each time interval
separately, {\tt mkgrmf} for the response files (only one file for all
time intervals). We excluded data around three absorption edges
because the effective area curve there is affected by uncertainties in
the model of the detector contaminant.  Data were excluded in the
following photon energy intervals: 0.52--0.55 (oxygen K edge),
0.68--0.73 (iron L edge) and 0.84--0.89 keV (neon K edge). Orders were
combined through {\tt add\_grating\_orders}. Despite the high count
rates during the burst, none of the {\it Chandra}\/ data are
significantly piled up, due to the small frame time and large
dispersion. The burst spectra were also independently extracted and
fit using {\tt idl} and {\tt isis}, confirming all spectral fit
results.

\begin{table}
\caption[]{Spectral parameters of fit to non-burst or persistent {\it
    Chandra}/{\it RXTE}\/ spectrum in 2.5 ks prior to
  burst.\label{tab}}
\begin{tabular}{ll}
\hline\hline
Parameter & Fitted value\\
\hline
$N_{\rm H}$ & $(0.88\pm0.04)\times10^{21}$~cm$^{-2}$ \\
disk black body k$T_{\rm in}$ & 0.591$\pm$0.006~keV \\
black body norm. ($R_{\rm in}^2/D^2_{\rm 10 kpc}$) & $(6.5\pm0.3)\times10^{2}$\\
power law photon index & 1.892$\pm$0.013\\
power law flux at 1 keV & 0.399$\pm$0.014 ph~s$^{-1}$cm$^{-2}$keV$^{-1}$ \\
gauss centroid & 5.12$\pm$0.09~keV\\
gauss sigma & 1.66$\pm$0.06~keV\\
gauss normalization & 0.0156$\pm$0.0012~phot~s$^{-1}$cm$^{-2}$\\
$\chi^2_\nu/\nu$ & 0.803/1530\\
0.5-20 keV flux & $(3.738\pm0.008)\times10^{-9}$~\ecs \\
0.5-60 keV abs. flux & $4.772\times10^{-9}$~\ecs \\
0.5-60 keV unabs. flux & $5.196\times10^{-9}$~\ecs \\
\hline\hline
\end{tabular}
\end{table}

{\it RXTE}/PCA spectra were extracted from event mode data using ftool
{\tt seextrct}, filtering out artificial 'clock events'. The PCA
spectral response was calculated with {\tt pcarsp} version 11.7.1. We
find that the background model for PCU1 underestimates the spectrum
beyond 30 keV, where no source photons are expected, by about 25\%. To
optimize the statistical quality and employ as many photons as
possible, we chose to exclude the bandpass beyond 20 keV rather than
to exclude PCU1 data. Below 20 keV the source photon spectrum is at
least four times brighter than the background spectrum, so that the
error is at most 3\% (at 20 keV) for the non-burst emission and less
for the burst emission. We verified the results with those obtained
after excluding PCU1 and do not find significantly different
results. Thus, we employ data of both active PCUs.  The dead time
fraction was calculated from standard-1 data following the
prescription at the {\it RXTE}\/ web
site\footnote{\url{http://heasarc.gsfc.nasa.gov/docs/xte/recipes/pca_deadtime.html}}
and was found to be up to 23.3\% at burst peak. This fraction was
taken into account when calculating energy fluxes. All spectral bins
up to 20~keV contained enough photons for Gaussian statistics to
apply.

\section{Spectral analysis}
\label{secspec}

{\sc XSpec} version 12.7.1b was employed as analysis tool.  {\it
  Chandra}\/ and {\it RXTE}\/ burst spectra were simultaneously
modeled when investigating the continuum. {\it RXTE}\/ data were
fitted between 3 and 20 keV and {\it Chandra}\/ data between 0.5 and 6
keV. All spectral bins were grouped so that each bin contains at least
15 photons. No background was subtracted from the {\it Chandra}\/
spectra, while the particle-induced plus cosmic diffuse background was
subtracted from the {\it RXTE}\/ spectra (employing {\tt
  pcabackest}). The {\it Chandra}\/ background is expected to be a
minor part (a few c~s$^{-1}$, see {\it Chandra}\/
website\footnote{\url{
    http://cxc.harvard.edu/contrib/maxim/bg/index.html}}) of the
non-burst emission (117~c~s$^{-1}$ on average).  We kept the
normalization factor of the {\it Chandra} data with respect to the
{\it RXTE} data to 1. When left free, the fit procedure finds a
minimum $\chi^2_\nu$ at a different normalization for every spectrum
by adjusting other spectral parameters. Furthermore, the improvement
in $\chi^2_\nu$ is never more than about 1\%. We conclude that the
normalization is ill constrained. Thus, fluxes are calibrated against
the PCA response.

\subsection{{\it Chandra}/{\it RXTE}\/ pre-burst spectrum}
\label{secnb}

A good representative of the non-burst emission during the burst may
be obtained from data during the 2.5 ks prior to the burst. The flux
remains approximately constant during that time and is similar to the
flux immediately after the burst (for a high-quality light curve from
{\it RXTE}\/ data, see Fig.~\ref{figlcxte}). Furthermore, there is
simultaneous {\it Chandra}\/ and {\it RXTE}\/ data for this interval.
We derived a satisfactory empirical model for these data consisting of
a disk black body \citep[e.g.,][]{mit84}, a power law and a gaussian
component, all absorbed by a single medium modeled according to the
prescription in \cite{wil00}. The Gaussian component only applies to
PCA data. It was also often needed in PCA spectra of other X-ray
bursts \citep{gal08} and may be related to the collection of
additional Galactic emission in the wide field of view of the
PCA. Table~\ref{tab} presents the best-fit model parameters.  The
value for $N_{\rm H}$ is within the range of values found previously
by, for instance, \cite{cac09} and \cite{pap09}.

\subsection{Burst spectrum}

\begin{figure}[t]
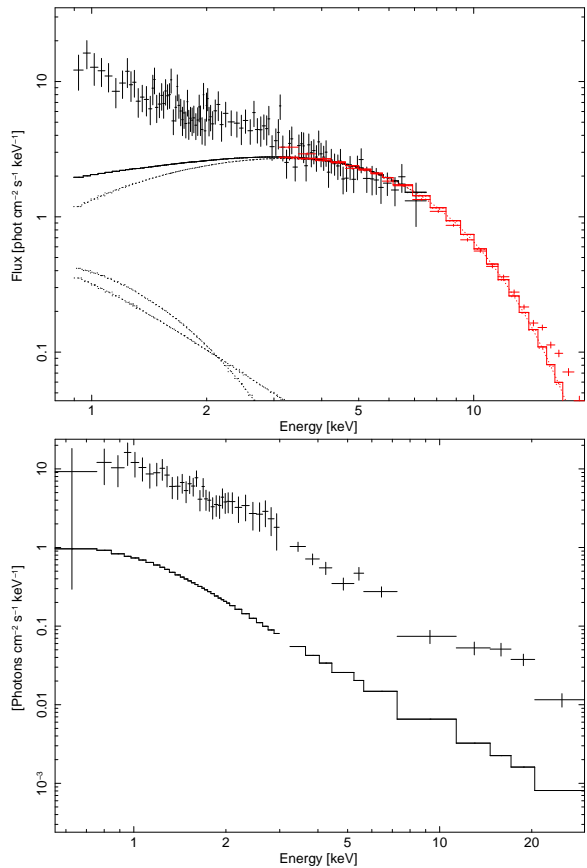

\centerline{\includegraphics[height=0.9\columnwidth,angle=270]{intzand_AA21056_f4a.ps}}
\centerline{\includegraphics[height=0.9\columnwidth,angle=270]{intzand_AA21056_f4b.ps}}
\caption{Deconvolved spectra for 2 to 5 s after burst onset. The top
  panel shows the {\it Chandra}\/ (black) and {\it RXTE}\/ (red)
  spectrum. Data points are indicated by crosses and the total model
  by the histogram. The upper dotted curve represents the black body
  component and the lower two curved lines comprise the model
  components for the pre-burst \sout{non-burst} radiation. The model
  is a bad fit ($\chi^2_\nu=3.505$ for $\nu=268$), but can be made
  consistent with the data by leaving free the normalization $f_{\rm
    a}$ of the non-burst model ($\chi^2_\nu=0.963$ for $\nu=267$). In
  the bottom panel the residual spectrum is plotted after subtraction
  of the fitted black body model (crosses). The model for the
  non-burst spectrum is plotted as a histogram. For clarity {\it
    Chandra} data is plotted up to 3 keV and {\it RXTE}\/ data beyond
  3 keV (only PCU2 to extend the energy range from 20 to 30 keV for
  visualization purposes).
\label{figc}}
\end{figure}

We first fitted the burst data with an absorbed black body model,
keeping the non-burst model fixed. This is the most common model
applied to X-ray burst spectra. This model does not provide a
statistically acceptable fit to the data for the first 20 s of the
burst, neither the combined data nor the {\it Chandra}\/ and {\it
  RXTE}\/ data alone. This is illustrated in Fig.~\ref{figc} (top
panel) which shows the data for 2 to 5 s after burst onset when the
temperature is varying by no more than 10\% (see below), along with
the best-fit model; the goodness of fit is $\chi^2_\nu=3.58$ with
$\nu=267$ degrees of freedom. There is a strong excess at low photon
energies, particularly in the {\it Chandra}\/ data. In fact, the {\it
  Chandra}\/ data alone are best fitted with a power law instead of a
Planck function, in contrast to the {\it RXTE}\/ data.

In order to find an explanation for the soft excess, we tested four
spectral models. We did this in two time intervals, because the model
needs to be applicable over all times. We chose time intervals 2-5 and
12-16 s, since during these times the spectrum does not vary much
while the statistical quality is good.  The first two models take into
account inelastic scattering of photons by electrons in either the NS
atmosphere or the accretion disk. The former was first put forward by
\cite{lon86}, further developed by \cite{mad04} and most recently
calculated by \cite{sul11,sul12}. Hot atmospheric electrons harden the
photons coming from below and increase the observed temperature to a
value that is one to two times larger than the effective
temperature. A soft excess remains \citep[see Fig.~7 in][]{sul11}. The
model's free parameters are the luminosity in terms of the Eddington
value and the NS radius. The fit of this model to the first data set
(see Table~\ref{tabspec}) is insufficient, but that may be expected
because model does not formally apply to radiation at the Eddington
limit. The fit to the second data set, at sub Edddington flux levels,
is not acceptable either.  The data have a broader spectrum than the
model.

\begin{table*}
\caption[]{Fit results when attempting four different models (columns)
  to account for the soft excess on 2 data sets (rows). First order
  LETGS spectra were combined and binned to 15 photons per bin. The
  figures show the deconvolved spectra (top panels; in units of
  phot~cm$^{-2}$s$^{-1}$keV$^{-1}$) and their relative deviation
  (i.e., data divided by the predicted model value; bottom panels)
  with respect to the models. Black curves are for LETGS and red for
  {\it RXTE} data. The y-axes for the bottom panels have identical
  scales.\label{tabspec}}
\begin{tabular}{|l|l|l|l|}
\hline
{\bf NS atmosphere model} & {\bf Reflection model} & {\bf Double black body} & {\bf '$f_{\rm a}$' model} \\
\hline
\multicolumn{4}{c}{ } \\
\hline
\multicolumn{4}{|c|}{\large\bf Time interval: 2-5 s ($\chi^2_\nu=3.505$ with $\nu=268$ for black body fit)} \\
\hline
$\;\;\chi^2_\nu=2.965$ ($\nu=268$)& $\;\;\chi^2_\nu=1.099$ ($\nu=266$)& $\;\;\chi^2_\nu=1.274$ ($\nu=266$) & $\;\;\chi^2_\nu=0.963$ ($\nu=267$) \\
$\;\;L/L_{\rm Edd}=0.501\pm0.004$ & $\;\;$log$\xi=3.60\pm0.03$      & $\;\;kT_1=0.620\pm0.017$            & $\;\;kT=1.985\pm0.009$\\
$\;\;R=16.49\pm0.09$~km         & $\;\;kT=2.27\pm0.03$             & $\;\;F_1=$18\%                    & $\;\;F_{\rm bb}$=63\% \\
                                & $\;\;R=10.0\pm12.8$              & $\;\;kT_2=2.082\pm0.013$            & $\;\;f_a=16.6\pm0.6$\\
                                & $\;\;(n_{\rm H}=10^{20}$~cm$^{-3}$)& $\;\;F_2=$82\%                    & $\;\;F_{\rm f_a}$=37\% \\
\includegraphics[height=0.49\columnwidth,angle=270,trim= 00 130 30 0,clip]{intzand_AA21056_t2a.ps} & \includegraphics[height=0.49\columnwidth,angle=270,trim= 00 130 30 0,clip]{intzand_AA21056_t2b.ps} & \includegraphics[height=0.49\columnwidth,angle=270,trim= 00 130 30 0,clip]{intzand_AA21056_t2c.ps} & \includegraphics[height=0.49\columnwidth,angle=270,trim= 00 130 30 0,clip]{intzand_AA21056_t2d.ps} \\
\hline
\multicolumn{4}{c}{ } \\
\hline
\multicolumn{4}{|c|}{\large\bf Time interval: 12-16 s ($\chi^2_\nu=1.792$ with $\nu=148$ for black body fit)} \\
\hline
$\;\;\chi^2_\nu=1.893$ ($\nu=148$)& $\;\;\chi^2_\nu=0.954$ ($\nu=146$) & $\;\;\chi^2_\nu=0.693$ ($\nu=146$) & $\;\;\chi^2_\nu=1.341$ ($\nu=147$) \\
$\;\;L/L_{\rm Edd}=0.293\pm0.002$ & $\;\;$log$\xi=3.61\pm0.14$       & $\;\;kT_1=1.593\pm0.046$            & $\;\;kT=1.790\pm0.007$\\
$\;\;R=13.76\pm0.08$~km         & $\;\;kT=1.84\pm0.02$              & $\;\;F_1=$73\%                    & $\;\;F_{\rm bb}$=80\% \\
                                & $\;\;R=0.38\pm0.24$               & $\;\;kT_2=2.82\pm0.34$            & $\;\;f_a=3.7\pm0.3$\\
                                & $\;\;(n_{\rm H}=10^{20}$~cm$^{-3}$) &$\;\;F_2=$27\%                     & $\;\;F_{\rm f_a}$=20\%\\
\includegraphics[height=0.49\columnwidth,angle=270,trim= 00 130 30 0,clip]{intzand_AA21056_t2e.ps} & \includegraphics[height=0.49\columnwidth,angle=270,trim= 00 130 30 0,clip]{intzand_AA21056_t2f.ps} & \includegraphics[height=0.49\columnwidth,angle=270,trim= 00 130 30 0,clip]{intzand_AA21056_t2g.ps} & \includegraphics[height=0.49\columnwidth,angle=270,trim= 00 130 30 0,clip]{intzand_AA21056_t2h.ps} \\
\hline
\end{tabular}
\end{table*}

Scattering by the accretion disk is calculated through the reflection
model by \cite{bal04}. In this model, black body radiation is assumed
to hit the accretion disk and instantly photo-ionize it. The disk is
modeled as a constant-density 1-dimensional slab. The radiation then
is reprocessed by the disk and re-emitted into the line of
sight. Above a few keV the reflected spectrum is very similar to the
black body spectrum. Below a few keV it shows a soft excess whose
detail depends on the level of ionization. The magnitude of the soft
excess is a strong function of the density of the disk \cite[see
  Fig.~4 in][]{bal04}. The total amount of reflection depends on the
observer's viewing angle of the accretion disk. The model's free
parameters are the black body temperature $kT$, the ionization
parameter log$\xi$ ($\xi$ in units of erg~cm~s$^{-1}$) and the
reflection fraction $R$. We first tested the disk model with density
$n_{\rm H}=10^{15}$ H-atoms cm$^{-3}$ on the 2-5 s data, but this
turns out to be unacceptable. A model with $n_{\rm H}=10^{20}$ H-atoms
cm$^{-3}$ performs better and is able to fit the soft excess
reasonably. The results are provided in Table~\ref{tabspec}. The
reflection fraction turns out to be very large (10) for the 2-5 s
data. It cannot be forced to smaller values by leaving free other
parameters such as $N_{\rm H}$. That may suggest that $n_{\rm H}$ is
even higher, although 10$^{23}$ cm$^{-3}$ is close to the maximum
value in standard accretion disk theory \citep{sha73}. The strong
coupling between $R$ and $n_{\rm H}$ makes it difficult to obtain a
good constraint on both, though. Unfortunately, due to increasing
importance of 3-body interactions, there is no model yet available for
higher $n_{\rm H}$ values.

The third model involves the addition of a second black body component
with a different temperature. This is based on the possibility that
there is exists a boundary layer between the accretion disk and the
neutron star where radiation is released due to friction between both
\citep{ino99,ino10}. Fits with this model are fairly good, see
Table~\ref{tabspec}. We checked whether the small residual soft excess
for the 2-5 s data is resolvable with a smaller value for $N_{\rm H}$
but it is not completely ($\chi^2_\nu=1.229$ with $\nu=266$ for
$N_{\rm H}=0$~cm$^{-2}$).

The fourth model is a straightforward variation of the initial model,
namely to leave free the normalization $f_{\rm a}$ (with respect to
the pre-burst value) of the persistent emission component listed in
Table~\ref{tab}. This simple model was recently employed by
\cite{wor12b} to model successfully 332 PCA-detected X-ray bursts with
photospheric expansion in the 3-20 keV band. This model performs at
least as well in terms of $\chi^2_\nu$ as the double black body and
reflection model, see Table~\ref{tabspec}, but on top of that it shows
the least amount of systematic trend in the residuals for both
spectra. Therefore, we chose to perform the full time-resolved
spectral analysis with this model and note that the results for the
primary black body component are similar as in the double black body
model. Henceforth, we will call this the '$f_a$' model.

\begin{figure}[t]
{\includegraphics[width=\columnwidth,angle=0]{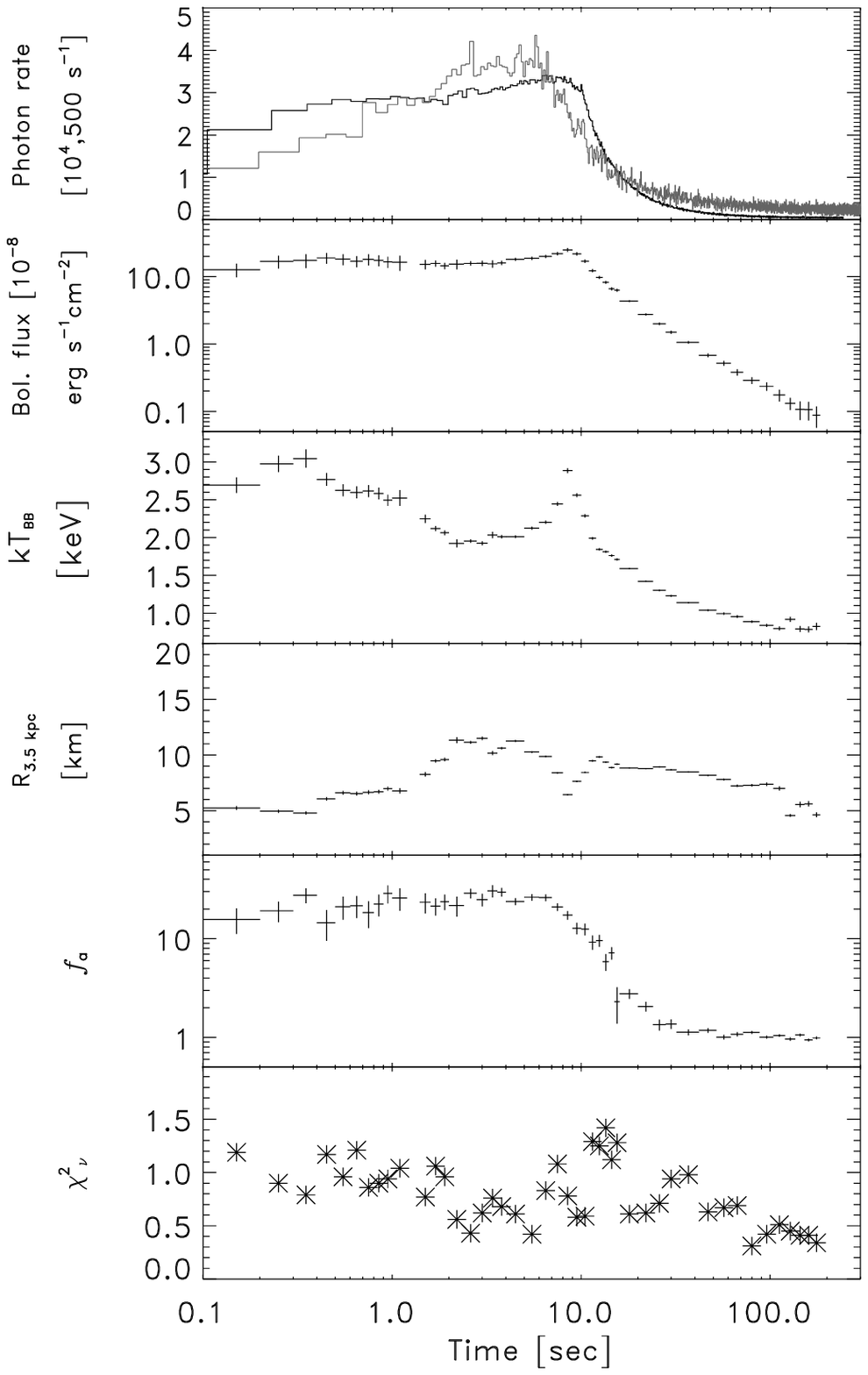}}
\caption{Time-resolved spectral analysis results for the combined {\it
    RXTE}/{\it Chandra}\/ data sets. The top panel shows the photon
  count rates in both instruments: gray for {\it Chandra}-LETGS (in
  units of 500 s$^{-1}$) and black for {\it RXTE}-PCA (in units of
  10$^4$ s$^{-1}$).
\label{figtrsp}}
\end{figure}

\begin{figure}
\vspace{0cm}
\centerline{\includegraphics[width=0.8\columnwidth,angle=0]{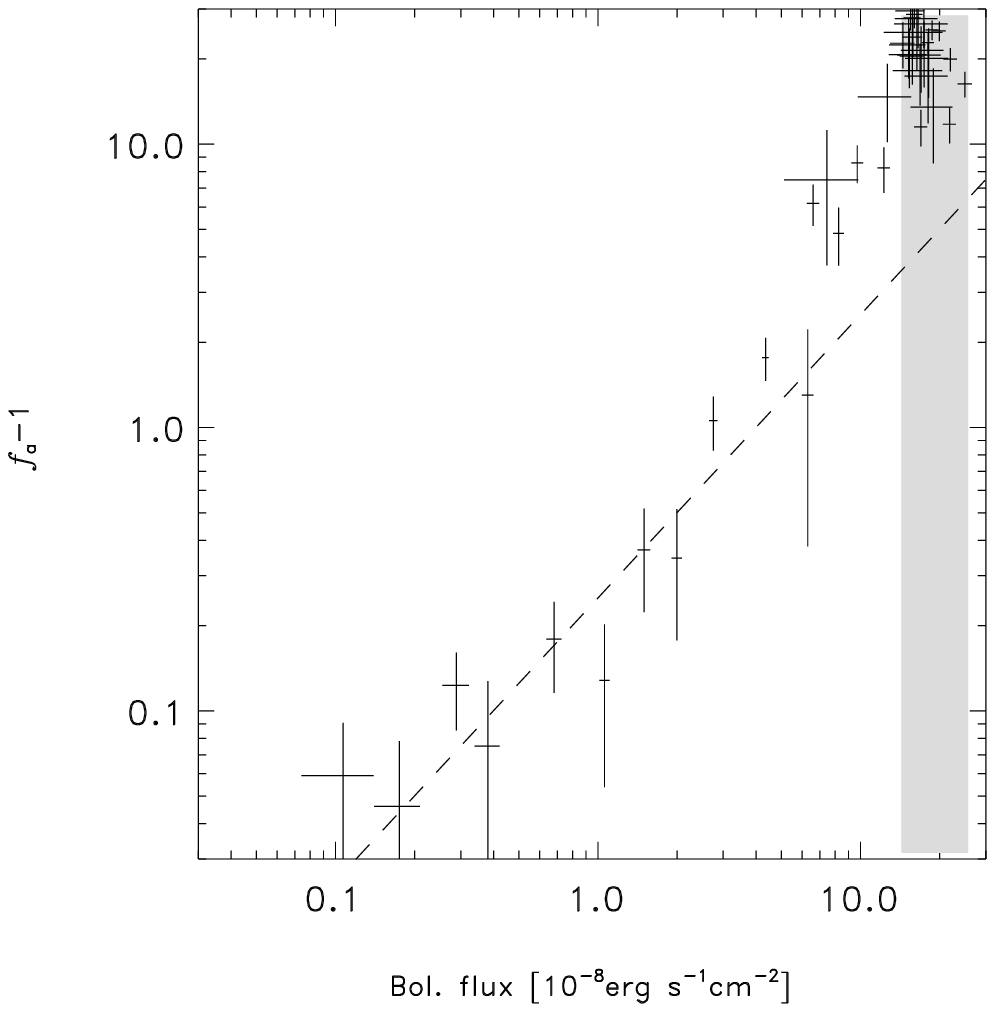}}
\caption{Persistent emission flux pre-factor $f_{\rm a}$ minus 1
  versus bolometric burst flux measurements (see also
  Fig.~\ref{figtrsp}).  The gray area indicates the region of the
  Eddington limit, between that for a hydrogen and a helium
  photosphere. The dashed line shows how the trend should go if the
  flux ratio of the increased persistent spectrum to burst would be
  constant.
\label{figcor}}
\end{figure}

The '$f_a$' model is illustrated in the bottom panel of
Fig.~\ref{figc} where the 2-5 s spectrum is shown after the fitted
black body model is subtracted. The model of the pre-burst spectrum is
also shown. Over the 0.5-30 keV range shown, the shape is the same to
a fairly accurate degree. To obtain a sense of how similar the
persistent spectrum during this interval is to that before the burst,
we left free in addition the power law and disk black body parameters.
The best-fit power law index becomes $1.83\pm0.05$ and the disk black
body $kT_{\rm in}=1.23\pm0.64$ ($\chi^2_\nu=0.701$ for
$\nu=263$). Therefore, it is somewhat harder.

We defined time bins covering 200~s from the burst onset, that were
sufficiently short to resolve the spectral variations, while
maintaining adequate signal-to-noise for precise spectral
parameters. The adopted binning, achieved via trial-and-error, totaled
47 approximately logarithmically-spaced bins between 0.1~s at burst
onset and 16~s in the tail.  All time bins except one bridge the {\it
  RXTE}\/ data gaps due to the full data buffer (see
Sect.~\ref{secobs}). In Fig.~\ref{figtrsp} we show the results of
fitting the '$f_a'$ model. The fits are excellent, except perhaps
between 10 and 20 s, see Sect.~\ref{sechr}. The soft excess remains
until 20-30 s after the burst onset. In Fig.~\ref{figcor} we plot
$f_{\rm a}-1$ against the burst bolometric flux. This shows that the
increase of the persistent spectrum is present for burst fluxes above
10\% of the peak flux and that its flux is not a simple constant
fraction of the thermonuclear burst flux.

The peak radius reached during the expansion interval was 10~km, a
factor of approximately 2 larger than the radius at touchdown. This
degree of expansion is quite modest; in the strongest radius expansion
bursts, the radius expands to at least $10^2$ times the NS radius. We
note that, similar to other bursts from this source, the radius
increases following touchdown up to a level comparable to the maximum
reached during the expansion. However, this increase during the burst
tail is usually attributed to a decrease in the spectral correction
factor as the flux decreases \citep[e.g.][]{sul11} so that the
photospheric radius during the expansion is significantly in excess of
the NS radius.

The burst reaches a peak flux of $(2.50\pm0.16)\times
10^{-7}$~\ecs\ (after accounting for the increased persistent
emission), and the bolometric burst fluence is $f_{\rm
  burst}=(3.41\pm0.04)\times10^{-6}$~erg~cm$^{-2}$ over 184~s.
Table~\ref{tab2} presents the complete set of derived burst
parameters. One can estimate the burst recurrence time according to
$t_{\rm recur}=\alpha \times f_{\rm burst} / F_{\rm pers}$, where
$\alpha\approx150$ \citep{gal06} and bolometric persistent flux
$F_{\rm pers}=6\times 10^{-9}$~\ecs.  This results in $t_{\rm
  recur}\approx80$~ks. This is similar to the maximum continuous data
set available around the burst (see Fig.~\ref{figlc}).  An interesting
number is the total number of bursts expected in the whole
outburst. If we assume a Crab spectrum for the whole outburst, the
total fluence under the estimated integral of the outburst light curve
(Fig.~\ref{figlc}) would be about $f_{\rm
  outburst}=2.9\times10^{-3}$~erg~cm$^{-2}$.  The total number of
expected bursts would then be $f_{\rm outburst}/\alpha f_{\rm burst}
\approx 6$. The net coverage of the {\it RXTE}\/ and {\it Chandra}\/
observations is only 2.5~d compared to roughly 25 d for the whole
outburst. Thus, it is not unexpected to have detected only one burst.

\subsection{Search for absorption edges}
\label{sechr}

\begin{figure}[t]
\centerline{\includegraphics[height=0.8\columnwidth,angle=270]{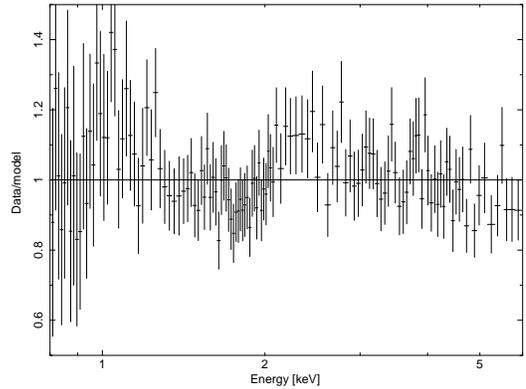}}
\caption{Combined and rebinned first-order LETGS spectrum of data from
  the first 20 s of the bursts, in terms of fractional residuals with
  respect to an empirical continuum fit.
\label{figedge}}
\end{figure}

\begin{figure}[t]
\centerline{
\includegraphics[width=0.9\columnwidth,angle=0]{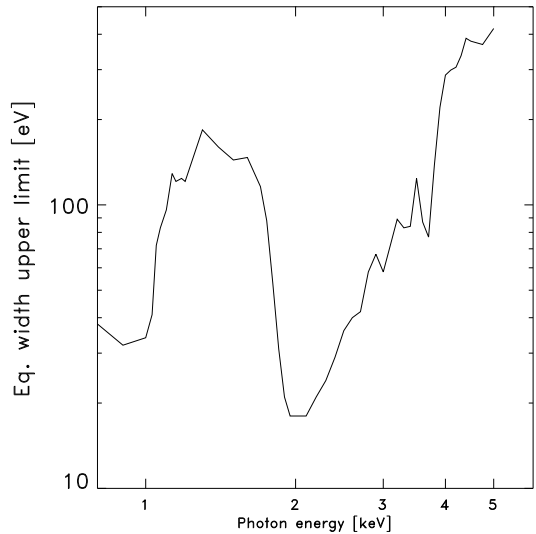}
}
\caption{For the 20 s exposure (see Fig.~\ref{figedge}), this graph
  shows the upper limits of an absorption edge as a function of photon
  energy.
\label{figewstraight}}
\end{figure}

We extracted full-resolution orders -1 and +1 of the LETGS data for
the first 20 s and fitted the spectrum after subtraction of the
pre-burst spectrum with a power law and show in Fig. \ref{figedge} the
deviations of the data with respect to the continuum model. The
goodness of fit is $\chi^2_\nu=0.684$ ($\nu=678$). There is no obvious
sought-after edge visible. We determined the 90\% confidence upper
limits on the equivalent width\footnote{we define the equivalent width
  as the integral over energy $E$ of \\$1-{\rm
    exp}(-\tau\left(\frac{E}{E_0}\right)^3)$ with $E_0$ the edge
  energy and $\tau$ its optical depth} of any absorption edge by
fixing a model for the edge at various photon energies, leaving free
the power law parameters and edge depth while fixing $N_{\rm H}$ and
determining the edge depths for which $\chi^2$ is 2.7 above the
minimum value. We find the upper limit to range between 20 eV at
1.8--2.4 keV to 200 eV at 1.3 and 3.3 keV up to 400 eV at 5.0 keV,
see Fig.~\ref{figewstraight}.

There is a broad dip at 1.8 keV which may arise from a broad
absorption line or edge feature.  Fitting it with an absorption line
yields a centroid energy of $1.77\pm0.04$~keV, a line width of
$\sigma=0.15\pm0.05$~keV and an equivalent width of $-56\pm15$ eV.
The goodness of fit is $\chi^2_\nu=0.624$ ($\nu=673$).  Fitting it
with an absorption edge yields an edge energy of 1.29$\pm$0.01 keV and
an optical depth of 0.29$\pm$0.06 with $\chi^2_\nu=0.648$
($\nu=676$). Adding a second edge does not really improve the fit,
with $\chi^2_\nu=0.641$ ($\nu=674$).

We searched for edges in smaller time frames and found none. The upper
limits are a factor of 4 to 5 worse at 1 s resolution than at 20 s.

In the {\it RXTE}\/ data we analyzed the spectra with the highest
$\chi^2_\nu$ values: between 6 and 8 s (this is near the touch-down
point) and between 12 and 16 s. These spectra show spectral deviations
which are reminiscent of the shape of an absorption edge with an
energy between 7 and 8 keV and an optical depth of 0.2-0.3 (equivalent
width 0.7-1.1 keV). Unfortunately, these edges are not covered by the
{\it Chandra}\/ data so we cannot get confirmation from there. We
repeated the time-resolved spectroscopic analysis illustrated in
Fig.~\ref{figtrsp}, but adding an absorption edge and find that the
data are consistent with the continuous presence of the above
mentioned edge, but that it is detectable only when the statistical
quality of the data are sufficient to allow detection which is between
roughly 6 and 40 s after burst onset.

\subsection{Search for spectral lines}
\label{seclines}

We searched for absorption and emission lines in the LETGS data during
five time intervals: 0-2 s during the burst rise, 2-5 s during the
first part of the peak, 5-9 s during the second part of the peak, 9-12
s during the first part of the tail and 12-16 s during the second part
of the tail. A 4.0$\sigma$ detection threshold was employed. This is
equivalent to one chance detection in all 20,000 trials. The number of
trials is determined as follows: 200 independent tests per 2-12 \AA\
spectrum with a LETGS resolution and binning of 0.05 \AA; a factor of
2 for five successive 2 times binned-up versions of the spectra; and a
factor of 5 for the five time intervals mentioned above. In order to
have a 10\% chance of a false positive (in either absorption or
emission), a one-tailed test would require a 1 in 20,000 chance for
detection in random data, which is about 4.0 sigma.  We also searched
in 4 different time intervals (0-7, 2-4, 4-6, and 7-30.5 s) that
overlap the above time intervals.

\begin{figure}[t]
\centerline{\includegraphics[width=1.\columnwidth,angle=0]{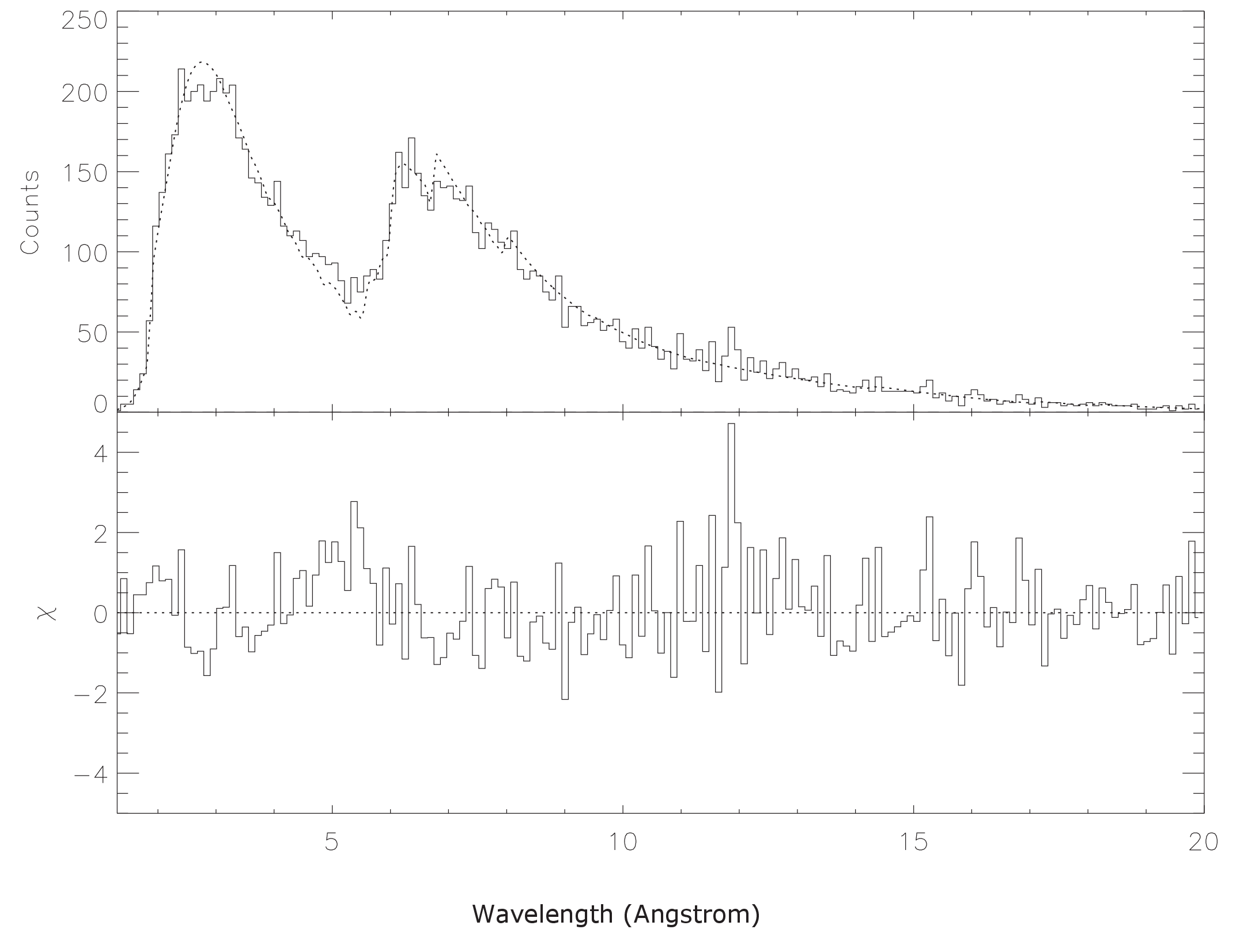}}
\caption{{\it Chandra}-LETGS burst spectrum between 0 and 7 s after
  burst onset, in wavelength domain. The histogram represents the data
  and the dotted line a continuum fit. Top panel is the spectrum,
  bottom panel the deviation with respect to the fitted model in units
  of $\sigma$. Note the feature at 11.9 \AA.
\label{figline}}
\end{figure}

The strongest candidate line that we find is an emission feature that
we find in the 0-7 s time interval at 11.9 \AA\ (or 1.04 keV), see
Fig.~\ref{figline}. This feature is 4.7$\sigma$ above the continuum in
a binning of 0.11 \AA. Given that we searched in additional time
intervals, the significance should be diminished to 4.6$\sigma$. The
equivalent width is about 4 \AA.

\section{Timing analysis}
\label{sectiming}

We performed a timing analysis of the {\it RXTE}\/ data of the burst,
by generating Fourier power spectra for 10-s data stretches every 1 s,
employing the event mode data with 125 $\mu$s resolution.  We limited
the analysis to channels 5--37 (2--10 keV), where the ratio of the
burst photon rate to the background rate is highest. We detect burst
oscillations with a burst fractional amplitude of 2--5\% rms in the
tail.  This value is calculated without taking into account the
contribution of the accretion powered pulsations that might still be
present at the time of the burst. This correction is relatively small
since the accretion powered pulsations have pulsed fractions of 3--4\%
rms prior and after the occurrence of the burst \citep[see for
  example][]{pat12b}. Furthermore the photon count rate detected when
burst oscillations are seen is between two and four times the count
rate received prior to the occurrence of the burst, meaning that the
oscillations seen are truly of nuclear origin (see e.g., Eq. 11 in
Watts et al. 2005). Similar considerations apply to the burst
oscillation phase, which are negligibly influenced by the presence of
accretion powered pulses. Burst oscillations were marginally detected
(between 2 and 3 sigma for 1~s long trains) at 400-401 Hz during the
burst rise with fractional rms amplitudes of up to 5\%. The formal
90\%-confidence upper limit on the fractional rms amplitude is
8\%. This is consistent with measurements of previous bursts from
\bron\ with RXTE \citep{chak03,bha06} where amplitudes were seen
between 5 and 25\% in 0.25~s long trains. Our data is less sensitive
because of the smaller number of active PCUs (2 instead of 4).

The time resolution of the ACIS-S CC mode data is 2.85 ms (351
Hz). This implies that the burst oscillation signal (401 Hz) cannot be
easily resolved in the {\it Chandra} data.  Simulations show that to
detect a sinusoidal pulsation at 401 Hz in a 1~s stretch of data with
a resolution of 2.85 ms requires a relative rms amplitude of at least
25\% to be significantly detected as an alias at 401-351=50 Hz which
is an order magnitude larger than ever detected for \bron.

\section{Discussion}
\label{secdiscuss}

The 2011 outburst of \bron\ is the seventh since its discovery in 1996
\citep{zan98}, and comes 3 years after the previous outburst.
\cite{gal08c} found that the onset time of the earlier outbursts
followed a quadratic relation with time, which predicted the time of
the previous outburst to within 11~d. However, the 2011 outburst
occurred 197~d earlier than predicted, suggesting that the long-term
mass transfer rate may be increasing.  We estimated the total fluence
for the outburst as $\approx 2.9\times10^{-3}$~erg~cm$^{-2}$, which is
significantly lower than for the first four outbursts \citep{gal06b}.
Thus, it seems likely that the long-term accretion rate remains
substantially below that of the first decade of observations.

\subsection{Peak flux and burst oscillations}

The thermonuclear burst detected from \bron\ on 2011 November 8 is the
brightest ever detected with {\it Chandra}, and the second brightest
with {\it RXTE}.  Within the 1907~ks exposure time on \bron\ during
the complete {\it RXTE}\/ mission, this is the ninth burst
\citep[e.g.,][]{gal08}. The second brightest burst has a peak flux of
$(2.40\pm0.03)\times 10^{-7}$~\ecs\ \citep[this is burst no. 4
  in][note that this publication lists peak fluxes before dead time
  correction]{gal08} that is only slightly fainter than the
brightest. Our burst is otherwise fairly ordinary for \bron, including
its timing behavior. \cite{gal06} find that bursts from \bron\ are due
to flashes in a hydrogen-depleted layer and burn primarily helium.

The Eddington limit of a solar-composition photosphere on the surface
of a 10~km radius NS of mass 1.4~M$_\odot$ at a distance of 3.5 kpc
corresponds to a flux of 1.4$\times10^{-7}$~\ecs, for a helium-rich
photosphere 2.6$\times10^{-7}$~\ecs. The peak flux for the burst from
\bron\ is within this range, clearly indicating that the energy
production rate due to the thermonuclear flash is consistent with the
Eddington rate. It is perhaps worth noting here that the distance
estimate of 3.5~kpc for \bron\ comes from a comparison of the measured
burst energetics and recurrence times with the predictions of
theoretical ignition models \citep{gal06}, not merely from the peak
flux of previous PRE bursts, as is the case for the majority of LMXB
distances.  While the time dependence of the black body temperature
and normalization is also characteristic for an Eddington-limited
burst, with the typical 'touch-down' peak at 10 s (see
Fig.~\ref{figtrsp}), the radius expansion is limited to at most a
factor of about 2. This is typical for moderate expansion
\citep{zan10}. There is no evidence for the expulsion of a shell, such
as a (short) precursor. Ergo, this is not a superexpansion burst (see
Sect.~\ref{secintro}).

The accretion flux prior to the burst translates to a 0.5-20 keV
luminosity of 4.5$\times10^{36}$~\lum\ for a distance of 3.5 kpc, or
64 times smaller than the burst peak in the same bandpass.

The burst oscillations detected in this burst are very similar to
those detected in the {\it RXTE}\/ data of previous bursts. The
amplitude compares well with the fractional rms amplitude of 2-4\% in
previous bursts \citep[][]{wat12,pat12a}.

\begin{table}[t]
\caption[]{Burst parameter values, extracted from {\it RXTE}\/ data alone for
  easy comparison with other bursts
  \citep[e.g.,][]{gal08}. Times are non-barycentered.\label{tab2}}
\begin{tabular}{ll}
\hline\hline
Parameter & Value \\
\hline\hline
Onset time & MJD \mjdstart\ (barycentered) \\
            & 2011 Nov 8 21:55:00.50 UTC \\
            & 563407175.90~s {\it RXTE} MET \\
            & 437176836.46~s {\it Chandra} MET  \\
Rise times & $2.5\pm0.1$~s (25 to 90\% of peak)\\
           & $6.3\pm0.1$~s (0 to 100\%)\\
           & $0.1\pm0.1$~s (0 to 70\%)\\
e-folding decay time & 12.0 s \\ 
Time scale (fluence/peak flux) & 12.9 s \\
Duration & $\approx$500 s (down to 0.03\% of peak flux)\\
Unabs. bol. peak flux & $(2.56\pm0.15)\times 10^{-7}$~\ecs \\
Unabs. bol. fluence & $(3.41\pm0.04)\times10^{-6}$~erg~cm$^{-2}$ \\
3-20 keV persistent flux & $1.72\times10^{-9}$~\ecs \\
0.5-60 keV unabsorbed flux & $6.1\times10^{-9}$~\ecs\\
\hline\hline
\end{tabular}
\end{table} 

\subsection{Broad-band spectrum}

The {\it Chandra}\/ data are interesting for the peculiar behavior of
the burst spectral continuum. As illustrated in Fig.~\ref{figc}, data
below 3-4 keV draw a rather different picture of the burst spectrum
than above. The sub-3 keV data reveal a strong soft excess above the
black body. This is not the first time that a soft excess is observed
in a burst spectrum, although it seems to be a rare phenomenon,
possibly because many bursters have too much interstellar absorption
to allow detection of a soft excess. We count about 250 X-ray bursts
that have been detected with sub 3 keV coverage by {\it XMM-Newton},
{\it Chandra}\/ and {\it Swift}-XRT
\citep[e.g.,][]{kon07,zan09,gal10}. Most of these do not show
photospheric expansion. Only \cite{asai06} and \cite{boi07} report on
strong soft excesses found in some bursts from the eclipsing LMXB
EXO~0748$-$676. They propose that this is due to a decrease in
absorption by the circumstellar medium arising from the
photo-ionization of that medium by the burst photons, thereby allowing
more accretion disk flux to reach the X-ray detector.  This
explanation may be plausible for a high-inclination system
($i=75^\circ$--83$^\circ$ for EXO~0748$-$676; \cite{par86}), but for
\bron\ the inclination angle is suggested to be low
\citep[$<67^\circ$;][]{del08}, so there is little local obscuring
material affecting the spectrum. We note that the two published {\it
  Chandra\/} HETGS grating observations of 33 X-ray bursts
\citep[including 4 with photospheric expansion;][]{tho05,gal10} could
have found soft excesses of similar magnitude as in \bron, had they
been present.

The net burst spectrum is not consistent with the NS atmosphere model
of \cite{sul11}. There is better agreement with the model for
reflection against the accretion disk and even better with the simple
double black body model. Mildly better overall is the agreement with
the '$f_a$' model, because this models appears to have the smallest
systematic trends in its residuals (see bottom panels of figures in
Table~\ref{tabspec}). In a study of 332 radius-expansion bursts seen
with {\it RXTE}, \cite{wor12b} modeled the 3-20 keV spectra in the same
'$f_a$' manner. They found as a general feature that $f_a$ is
significantly in excess of 1, with the highest values arising from the
bursts observed at the lowest pre-burst fluxes. Our combined {\it
  Chandra-RXTE} analysis of one burst from \bron\ is consistent with
those results and shows additionally that this behavior extends to
lower energies, giving rise to a large flux excess below 3 keV for a
system with low $N_{\rm H}$ (see Fig.~\ref{figc}). However, we note
that the difference in performance with other models is not much
better. It is in the spirit of Occam's Razor (with only 1 additional
free parameter) that we provide the full time-resolved spectroscopic
analysis in terms of the simple '$f_a$' model, but other models are
almost equally justified.

\cite{wor12b} suggest that the increase of the non-burst emission is
caused by the Poynting-Robertson (PR) effect, which removes angular
momentum from the inner accretion disk due to radiation drag, thereby
increasing the accretion rate during the burst.  If the accretion rate
in \bron\ increased by a factor of 20 during the burst, based on our
maximum best-fit value of $f_a$, it would come close to 50\% of the
Eddington limit. Significant changes in the persistent spectral state
might be expected, analogous to the so-called spectral state changes
of the persistent emission that are attributed to changes in the
geometry of the accretion disk \citep[e.g.,][and references
  therein]{done07}. Since the persistent spectrum of \bron\ at the time
of the burst is quite hard, a softening of the spectrum seems most
likely. However, it is not obvious that a typical hard-to-soft
spectral state change can take place in response to an increase in
accretion rate sustained for $\lesssim 20$~s, or even if the effects
of PR drag would produce the same observable effect. Our data suggests
that, if there is a change, the persistent spectrum instead becomes
harder during the burst.

We note also that if the accretion rate increases by a factor of 20
over 10~s, the additional matter must be that which would otherwise
have been accreted over the next 200~s, and one would thus expect a
suppression of the accretion rate following the burst. This would be
the hallmark of the PR effect taking place. We found no evidence of
$f_a$ values smaller than 1, out to 500~s after the burst. However,
the recuperation time of the disk may be longer if the disk viscosity
is relatively small \citep{wal92}, so that the suppressed accretion
rate would be spread out over a longer interval, and hence
undetectable.

The pre-burst spectrum likely includes a component from inverse
Compton scattering of undetectable soft photons (with energies $<
0.5$~keV for this burst) by an accretion disk corona
\citep[ADC;][]{whi82,beg83a,beg83b}. An increase in the persistent
spectrum would therefore require either a larger ADC, or an increase
of soft seed photons into the ADC. The burst cannot produce a larger
ADC through photoionization, as the temperature of the plasma can only
be raised to its Compton temperature which is $\approx kT$ for a
blackbody spectrum. This temperature is much lower than the $> 10$~keV
temperature required to explain the high-energy extent of the
persistent emission. Likewise, since the local dissipation in the
accretion disk scales as $R^{-3}$ (where $R$ is the radius along the
disk), expansion of the ADC outwards through a wind or outflow would
not be able to explain the significant increase in the ADC
emission. This constraint could be circumvented if the disk and ADC
were expanded \emph{inwards} by the burst, as through
Poynting-Roberston drag (see discussion above).

Perhaps the simplest explanation for an increase in the persistent
spectrum during the burst is a larger influx of seed photons produced
by reprocessing in the accretion disk. As discussed in Sect. 4.2, high
density reflection models were able to explain a large fraction of the
strong soft excess seen early on in the burst. This emission, which
extends below $0.5$~keV, is dominated by thermal bremsstrahlung
produced by the X-ray heated outer layers of the disk
\citep{bal04,bal05}, and is a natural source of additional seed
photons to the corona. This explanation has several advantages: first,
the absorption and re-emission of the burst will all occur on not much
more than twice the light-crossing time of the system, and would
explain why the enhanced persistent emission is seen even in the
earliest phases of the burst. Similarly, Figs. \ref{figtrsp} and
\ref{figcor} illustrate that the increase in the persistent flux falls
rapidly as the illuminating blackbody fades, which would be a natural
consequence of less reprocessing in the inner disk. Finally, if
reprocessing accounts for most of the soft excess, then this reduces
the values of $f_a$ required to fit the spectra. Nevertheless, this
idea remains just informed speculation, and further study of the
broadband spectra of Type 1 X-ray bursts, as well as a concentrated
theoretical effort into understanding the interaction of a burst with
the surrounding accretion flow, is needed to understand the physical
processes at work in these environments.

\begin{figure}[t]
\centerline{
  \includegraphics[width=0.9\columnwidth,angle=0]{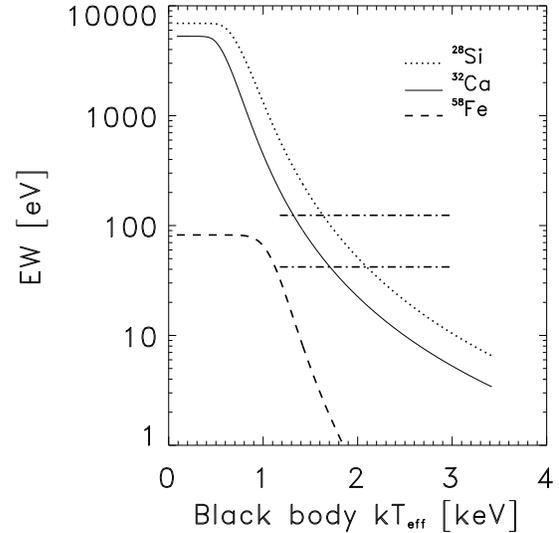} }
\caption{For the 20 s exposure (see Figs.~\ref{figedge} and
  \ref{figewstraight}), this graph shows the upper limits of an
  absorption edge at 2.67 keV (lower horizontal dash-dotted line) and
  3.48 keV (upper horizontal line), compared to the predictions for
  the K-edge of hydrogen-like $^{28}$Si (dotted curve), $^{32}$Ca
  (solid curve) and $^{56}$Fe (dashed curve), for abundances as
  determined from the He0.1 model (pure helium ignition at 0.1
  $\dot{M}_{\rm Edd}$) in \cite{wei06}. The range where the upper
  limits apply are set by the range of black body color temperatures
  measured (2-3 keV) multiplied with a maximum possible color
  correction of 0.6 to arrive at effective temperatures
  \citep[e.g.,][]{sul11}. The upper limits would be 3 to 4 times
  better if the edge energy would be gravitationally redshifted by a
  factor of 1.31. They are 4.5 times worse for an exposure time of 1
  s.
\label{figew}}
\end{figure}

\subsection{Evidence for spectral features}

The high spectral resolution data provided by the {\it Chandra}/LETGS
show no evidence for the sought-after absorption edge with upper
limits in the equivalent width of at best 20~eV. The residuals between
1 and 2~keV with respect to a continuum fit (Fig.~\ref{figedge}) are
not clearly identifiable with a single absorption edge or two. A
single edge would have an energy of 1.29~keV.  For a gravitational
redshift anywhere between $1+z=1$ for far away from the NS and
$1+z=1.31$ on the surface (for a NS with mass $1.4\ M_\odot$ and
radius $10$~km), the expected range of edge energies is
1.3--1.8~keV. This is not identifiable with a K or L-edge of any
likely element. The redshifted K-edge of hydrogen-like Si comes
closest at 2.0 keV. Only K-edge energies of Mg and Al and L-edge
energies of Ga, Ge, As, Se and Br would fit. Also, the residuals in
this energy range are likely to be affected by a sharp change in the
effective area arising from the K-edge of the neutral silicon CCD
material.

For a 20~s exposure, the upper limit on the equivalent width ranges
between 42~eV at the Si K-edge (2.67~keV) and 124 eV at the Ca K-edge
(3.48~keV). These are the two most abundant elements expected for pure
helium ignition according to \cite{wei06} with the largest predicted
absorption edges. These upper limits are compared with predicted
values in Fig.~\ref{figew}. The predictions are for ignition in a pure
helium layer like for \bron\ \citep{gal06}. The predictions are at
(for Si) or below the measured upper limits. Thus, these measurements
are consistent with predictions. We note that the tentative edge
detected in the {\it RXTE}\/ data at about 7.9 keV has an equivalent
width of 0.7-1.1 keV and is, thus, much higher than the prediction for
the Fe-K edge. On the other hand, the picture that this edge draws is
not consistent with the expectation. The edge energy and optical depth
are consistent with being constant while the temperature of the
photo-ionizing radiation field changes between 3 and 1 keV. This is
different from the features seen by \cite{zan10} and suggests one
should be careful in associating this feature with matter close to the
NS.

We find marginal evidence for an emission line at 11.9 \AA\ with an
equivalent width of 4 \AA. This wavelength resides in the range of
rest wavelengths of a number of Ne lines and is slightly blue-shifted
from the Fe L line complex. We believe the line is too marginal to
draw any further physical conclusions. However, it is interesting to
note that recently a strong emission line has been detected
\citep{deg13} at about the same energy in a long superexpansion burst,
with an equivalent width that is about 3 times smaller than the
feature we see, but for a much longer duration (10 min).

Although this burst is Eddington limited, it does not exhibit strong
photospheric expansion. Such expansion will probe the predictions at
lower black body temperatures in Fig.~\ref{figew}. Thus, the
sensitivity would be better. We suggest to focus future searches for
absorption edges on burst with those, ergo the 'superexpansion'
bursts, since these have not yet been detected at high spectral
resolution. The most opportune sources to search in are 4U 1820-30 (in
globular cluster NGC 6624) and 4U 1724-30 (in globular cluster Terzan
2), because they show bursts every few hours to days (4U 1820-30 in
the low/hard state) with at least 25\% of the time superexpansion
\citep[]{zan10,zan12}.

\section{Conclusion}

A simultaneous {\it Chandra}-LETGS/{\it RXTE}\/ detection of a very
bright thermonuclear X-ray burst with moderate photospheric expansion
and low $N_{\rm H}$ presents a unique opportunity to study the 0.5-30
keV spectrum of such an event, with large sensitivity and high
spectral resolution below 6 keV. We find that
\begin{list}{$\bullet$}{\leftmargin=0.4cm \itemsep=0cm \parsep=0cm \topsep=0cm}
\item the spectrum show a strong deviation from black body radiation,
  particularly at low energies where it exhibits a soft excess that
  exceeds the black body by an order of magnitude. The non-Planckian
  component has a similar spectral shape as the pre-burst spectrum
  that is powered by accretion, suggesting that the accretion flow
  changes during the burst, probably due to the near-Eddington flux
  from the burst.  Exactly how that happens is difficult to determine,
  mostly because there is no detailed model yet available for the
  dynamic and radiative interaction of a burst with the pre-existing
  accretion flow, but most likely it is related to an increase of seed
  photons in the ADC. Future work should include a hydrodynamic model
  of the interaction between X-ray bursts and accretion disks and a
  comprehensive of study of $f_{\rm a}$ values over bursts with and
  without photospheric expansion;
\item no unambiguous spectral features are detected. The upper limits
  for absorption edges are, however, consistent with the theoretically
  predicted equivalent widths due to ejected ashes.  It is expected
  that edges are deeper for bursts with photospheric
  superexpansion. Therefore, future searches with {\it Chandra}\/
  should preferably concentrate on those. This is also underlined by
  the recent finding by \cite{deg13}. Regarding absorption edges in
  bursts with moderate photospheric expansion, investigating
  superbursts with current grating spectrometers would be interesting
  and, in the farther future, the proposed LOFT mission with a 10
  m$^2$ silicon drift detector \citep{fer12} has the right combination
  of sensitivity and spectral resolution to make progress.
\end{list}

\acknowledgements

We thank Valery Suleimanov, Juri Poutanen, Tullio Bagnoli, Hauke
Worpel, Daniel Haas, David Huenemoerder for useful discussions, and
Suleimanov and Poutanen for making available their model in a form
that is suitable for use in {\sc XSpec}. We are grateful to Harvey
Tananbaum, Nancy Wolk, Jeremy Drake and the Chandra team for their
support of this quick (3.5~d turn around) TOO. JZ and DG acknowledge
ISSI for the hospitality in Bern where part of this work was
performed. This research has made use of data and software, in the
application package {\tt ciao}, provided by the Chandra X-ray Center
(CXC), {\it RXTE} data provided by the {\it RXTE} Guest Observer
Facility, {\it MAXI}\/ data provided by RIKEN, JAXA and the {\it
  MAXI}\/ team, and {\it Swift}/BAT data provided by the {\it
  Swift}/BAT team.

\bibliographystyle{aa} \bibliography{intzand_AA21056}

\end{document}